\documentstyle[prl,aps,epsfig]{revtex}
\tightenlines
\newcommand{\bea}{\begin{eqnarray}}
\newcommand{\eea}{\end{eqnarray}}
\newcommand{\beq}{\begin{equation}}
\newcommand{\eeq}{\end{equation}}
\newcommand{\bay}{\begin{array}}
\newcommand{\eay}{\end{array}}
\begin{document}

\twocolumn[\hsize\textwidth\columnwidth\hsize\csname@twocolumnfalse\endcsname


\title{Measuring the Photon Polarization in $B\to K\pi\pi\gamma$}

\author{Michael Gronau
and Yuval Grossman}
\address{Department of Physics, Technion-Israel Institute of Technology \\
Technion City, 32000 Haifa, Israel}

\author{Dan Pirjol}
\address{Department of Physics, University of California at San Diego \\
9500 Gilman Drive, La Jolla, CA 92093} 

\author{Anders Ryd}
\address{High Energy Physics, California Institute of Technology \\
Pasadena, CA 91125} 
\date{\today}
\maketitle
\widetext
\begin{abstract}
We propose a way of measuring the photon polarization in radiative $B$ decays 
into $K$ resonance states decaying to $K\pi\pi$, which can test the Standard
Model and probe new physics. The photon polarization is shown to be measured 
by the up-down asymmetry of the photon direction relative to the $K\pi\pi$ 
decay plane in the $K$ resonance rest frame. 
The integrated asymmetry in $K_1(1400)\to K\pi\pi$, calculated 
to be $0.34\pm 0.05$ in the Standard Model, is measurable at currently 
operating $B$ factories.
\end{abstract}
\bigskip

]\narrowtext

The Standard Model (SM) predicts that photons emitted in rare $b\to s\gamma$
decays are left-handed \cite{Hurth}, up to small corrections of order
$m_s/m_b$, while being right-handed in $\bar b\to \bar s\gamma$.  
This feature is common to inclusive and exclusive radiative decays,
also when including long-distance effects in the latter case \cite{GrinPi}.
While measurements
of the inclusive rate agree reasonably well with SM calculations \cite{Hurth},
no evidence exists for the helicity of the photons in these decays. In several
models beyond the SM the photon in $b\to s\gamma$ acquires an appreciable
right-handed component
due to the exchange of a heavy fermion in the electroweak loop process.
For instance, in $SU(2)_L\times SU(2)_R\times U(1)$
left-right symmetric models \cite{LRM} this component may be comparable in
magnitude to the left-handed component, without affecting the
SM prediction for the inclusive radiative decay rate. An independent
measurement of the photon helicity is therefore of interest.

Several strategies have been proposed to look for signals of physics beyond 
the SM through helicity effects in $B \to X_s \gamma$.
In one method the photon helicity is probed through mixing-induced CP
asymmetries \cite{AGS}. In two other schemes one studies angular distributions
in radiative decays of $\Lambda_b$ baryons \cite{MaRe,HK} and in
$B\to \gamma (\to e^+e^-) K^* (\to K\pi)$ \cite{Kim,GrPi}. 
The methods using $B$ mesons are sensitive to interference between amplitudes 
involving photons with left and right-handed polarization. In the SM the 
interference is at a level of a few percent, and these methods become 
unfeasible at present $B$ factories also for larger interference due to 
insufficient luminosities. The 
methods using $\Lambda_b$ decays, measuring directly the photon polarization, 
rely on future hadron colliders or on extremely high 
luminosity $e^+e^-~Z$ factories.

In the present Letter we propose to measure the photon polarization in
exclusive radiative $B$ decays to kaon resonance states, $B\to K_{\rm res}
\gamma$. We will study in particular decays into an axial-vector meson, 
$K_1(1400)$, and into a tensor meson, $K^*_2(1430)$. This measurement will 
be shown to be feasible at currently operating $B$ factories.
An earlier suggestion to look for parity violation in $B\to K_1(1400)\gamma$ 
was made in \cite{Weinstein}. 
Radiative decays into $K^*_2(1430)$ were
observed both by the CLEO \cite{CLEO} and Belle \cite{Belle}
collaborations with branching ratios around $10^{-5}$.
In these experiments $K^*_2$ states were identified through the $K\pi$ decay
mode. $K_1$ states, which do not decay in this mode, are expected to be
observed in the $K\pi\pi$ channel.
As we will argue below, in order to probe the photon
helicity, one must study excited kaon decays into final states involving 
at least three particles. 

Let us explain first the necessary conditions for a theoretically
clean measurement of the photon helicity in radiative $B$ decays from
recoil hadron distributions. Since the photon helicity is odd under
parity, and since one only measures the momenta of final decay
products, spin information cannot be obtained from two body decays of
the excited kaon. It requires at least a three body decay in which one
can form a parity-odd triple product $\vec p_{\gamma}\cdot (\vec
p_1\times \vec p_2)$. Here $\vec p_{\gamma}$ is the photon momentum,
and $\vec p_1,~\vec p_2$ are two of the final hadron momenta, all
measured in the $K$-resonance rest frame. The average value of the
triple product has one sign for a left-handed photon and an opposite
sign for a right-handed photon.

The above correlation is, however, also T-odd. In order not to violate
time-reversal in the excited kaon decay, the decay amplitude must involve
nontrivial final state interactions. Usually this poses the difficulty of 
introducing an unknown final state phase.
In order to have a measurement which can be cleanly interpreted in terms of 
the photon helicity, this phase difference must be calculable. This is the 
case in $K_{\rm res}\to K^*\pi \to K\pi\pi$, where two isospin-related
$K^*(892)$ resonance
amplitudes interfere. Parametrizing resonance amplitudes in terms of 
Breit-Wigner forms, known to be a very good approximation for the narrow 
$K^*$, yields a calculable strong phase. In this respect,
this method is similar to measuring the $\tau$ neutrino helicity in
$\tau \to a_1\nu_{\tau}$, where the corresponding phase-difference
is calculable in terms of the two interfering  $a_1 \to \rho\pi$ amplitudes
\cite{Kuhn,ARGUS}.

Considering cascade decays of $\bar B(b\bar q)~(q=u,d),~\bar B \to \bar 
K_{\rm res}\gamma \to \bar K\pi\pi\gamma$, we 
denote weak $\bar B\to \bar K_{\rm res}\gamma$ amplitudes involving left and 
right-handed photons by $c_L$ and $c_R$, and corresponding strong
$\bar K_{\rm res}$ decay amplitudes by ${\cal M}_L$ and ${\cal M}_R$, 
respectively. 
Amplitudes involving left and right-handed photons do not interfere
since {\it in principle} the photon polarization is measurable. Therefore,
\bea\label{A^2}
&& |A(\bar B\to \bar K_{\rm res}\gamma,~\bar K_{\rm res}\to \bar K\pi\pi)|^2 =
\nonumber \\ &&\qquad \qquad \qquad
|c_L|^2|{\cal M}_L|^2 + |c_R|^2|{\cal M}_R|^2\,. 
\eea
In the SM the photon in $\bar B$ decays is dominantly left-handed, 
$|c_R|^2 \ll |c_L|^2$. The corresponding $B$ decay amplitudes 
obey a reversed hierarchy implying a right-handed photon.
We denote the photon polarization by $\lambda_\gamma$,
\beq\label{lambda}
\lambda_{\gamma}\equiv {|c_R|^2 -|c_L|^2\over |c_R|^2 +|c_L|^2}~,
\eeq
such that in the SM $\lambda_{\gamma}\approx 1$ holds for radiative $B$ decays,
while $\lambda_{\gamma}\approx -1$ applies to $\bar B$ decays.

The weak amplitudes $c_{R,L}$ are given by $c_{R,L} = 
g_+^{K_{\rm res}}(0)C_{7R,L}$,
where $g_+^{K_{\rm res}}(0)$ are hadronic form factors at $q^2=0$, which 
have already been computed using several models \cite{gK}. (For most part, we 
will not rely on these calculations).
$C_{7R,L}$ are Wilson coefficients appearing in the 
effective weak radiative Hamiltonian
\bea
{\cal H}_{\rm rad} &=& -\frac{4G_F}{\sqrt2} V_{tb} V_{ts}^*\left( C_{7R}{\cal
O}_{7R} +  C_{7L}{\cal O}_{7L}\right)\,, \nonumber \\ 
{\cal O}_{7L,R} &=& \frac{e}{16\pi^2} m_b
\bar s\sigma_{\mu\nu}\frac{1 \pm \gamma_5}{2} bF^{\mu\nu}~.
\eea
Since the form factors $g_+^{K_{\rm res}}$ are common to $c_L$ and $c_R$, a
measurement of the ratio $c_R/c_L$ can be translated into
information about the underlying new physics entering the Wilson coefficients.

We now describe details of the method based on the decays
$B\to K_1\gamma$, beginning with formalism and ending with an 
estimate demonstrating the high sensitivity of the measurement to the 
photon polarization. We compare this sensitivity with the one using $K\pi\pi$ 
decays of $K^*_2$.

The decay processes $K_1(1400)\to K\pi\pi$ are dominated by 
$K^*(892)\pi$, with a branching ratio of $94 \pm 6\%$ \cite{PDG}. A smaller 
branching ratio into $\rho K$, $3  \pm 3 \%$ \cite{PDG}, will be neglected at 
this point,
and will be considered later on in order to estimate an uncertainty. We will 
study the modes
\bea\label{chain}
K^+_1&\to&
\left\{
\begin{array}{c}
 K^{*+}\pi^0 \\
 K^{*0} \pi^+ 
\end{array}
\right\} \to K^0 \pi^+ \pi^0\,,\nonumber \\
K^0_1 &\to&
\left\{
\begin{array}{c}
K^{*+}\pi^- \\
K^{*0} \pi^0 
\end{array}
\right\} \to K^+ \pi^- \pi^0~.
\eea

The decay amplitude of $K_1(1400)\to K^*(892)\pi$ can be written in terms of 
two invariant amplitudes
\beq\label{M1}
{\cal M}_1 = A(\varepsilon\cdot\varepsilon'^*) + 
B(\varepsilon\cdot p')(\varepsilon'^*\cdot p)~,
\eeq
where $\varepsilon,~p$ and $\varepsilon',~p'$ are the polarization vectors and 
momenta of the $K_1$ and $K^*$, respectively. This amplitude is a mixture of $S$ 
and $D$ waves. The $D/S$ ratio of widths and the phase difference between the 
two partial wave amplitudes were measured in \cite{Daum}, $|A_D/A_S|^2 = 0.04 
\pm 0.01$  and ${\rm arg}(A_D/A_S) \equiv \delta_{D}-\delta_S = 
(260 \pm 20)^\circ$, respectively.
The relation between the invariant amplitudes and the partial wave amplitudes
can be shown to be given by \cite{Isgur}
\bea\label{SD}
A &=& A_S + \frac{1}{\sqrt{2}}A_D~,\\
B &=& \left [-(1 - \frac{m_{K^*}}{E_{K^*}})A_S
-(1 + 2\frac{m_{K^*}}{E_{K^*}})\frac{1}{\sqrt{2}}A_D\right ] \frac{E_{K^*}}
{M_{K_1}{\vec p}^{\,2}_{K^*}}~,\nonumber  
\eea
where the $K^*$ energy and momentum are given in the $K_1$ rest frame.  
The amplitude (\ref{M1}) must be convoluted with the amplitude for $K^*\to K\pi$
which is proportional to $\varepsilon'\cdot(p_\pi - p_K)$. Isospin symmetry 
implies that the two $K^*$  contributions to the processes (\ref{chain}) are 
antisymmetric under the exchange of the two pion momenta. 

Denoting the momentum of $K_1$, the two pion momenta and the kaon momentum by 
$p,~p_1,~p_2$ and $p_3$, respectively, we find the amplitude of (\ref{chain}),
\beq\label{J}
{\cal M} = \varepsilon^{\mu}\,J_\mu~,~~
J_\mu = C(s_{13},s_{23})p_{1\mu} - (p_1 \leftrightarrow p_2)~,
\eeq
where
\bea
&& C(s_{13},s_{23}) \propto  
A \left[ \left(1 - \frac{m_K^2-m_{\pi}^2}{m_{K^*}^2}\right)
B^{K^*}_{23} - 2B^{K^*}_{13}\right] \\
&&+
B \left[\left(1 - \frac{m_K^2-m_{\pi}^2}{m_{K^*}^2}\right)
(p\cdot p_1 - m_\pi^2) - 2p_1\cdot p_2\right] B^{K^*}_{23}~, \nonumber 
\eea
and $B^{K^*}_{ij}$ is a Breit-Wigner form,
\bea
&&~~B^{K^*}_{ij} = \left( s_{ij} - m_{K^*}^2 - im_{K^*} 
\Gamma_{K^*}\right)^{-1}~,\nonumber \\
&&~~s_{ij} = (p_i + p_j)^2~.
\eea
$p\cdot p_1$ and $p_1\cdot p_2$ can be written in terms of $s_{13}$ 
and $s_{23}$.
Using (\ref{SD}), one obtains
\bea
&& C(s_{13},s_{23}) \propto
\left[ \left(1 - \frac{m_K^2-m_{\pi}^2}{m_{K^*}^2}\right)
B^{K^*}_{23} - 2B^{K^*}_{13}\right] \\
&&+
\kappa \left[\left(1 - \frac{m_K^2-m_{\pi}^2}{m_{K^*}^2}\right)
(p\cdot p_1 - m_\pi^2) - 2p_1\cdot p_2\right] B^{K^*}_{23}~,\nonumber 
\eea
where $\kappa = B/A = -[0.38 + 8.66|A_D/A_S|e^{i(\delta_D-\delta_S)}]
[1 + 0.71|A_D/A_S|e^{i(\delta_D-\delta_S)}]^{-1}{\rm GeV}^{-2}$.

Let us express the amplitudes ${\cal M}_{L,R}$ in the rest frame of
the $K_1$. The polarization vectors corresponding to right and 
left-handed $K_1$ of helicity $\pm 1$, $\varepsilon_{\pm 1}^{\mu}$, 
are defined in this frame by $\varepsilon^0_{\pm 1}=0$, and 
$\vec \varepsilon_{\pm 1} = \mp \frac{1}{\sqrt2} (\hat e_x \pm i\hat e_y)$. 
The two unit vectors $\hat e_x$ and $\hat e_y$ are perpendicular to $\hat e_z 
= -\hat p_\gamma$, which points along a direction opposite to the photon 
(or $B$) momentum. 
Denoting by $\theta$ the angle between the normal to the decay plane, $\hat n 
\equiv (\vec p_1\times \vec p_2)/|(\vec p_1\times \vec p_2)|$, and the 
direction opposite to the photon, $\cos\theta = \hat n \cdot \hat e_z$, 
one finds
\beq\label{M2}
{\cal M}_{R,L} \propto \frac{1}{\sqrt 2} \left ( \mp J_x -i\cos\theta J_{y'}
\right )~, 
\eeq
where $x,~y'$ and $\hat n$ form a set of orthogonal axes. (We choose these 
axes such that the plane perpendicular to the photon direction and the decay 
plane intersect on the $x$ axis.) 

Squaring the amplitudes and integrating over a common rotation angle $\phi$
of $\vec p_1$ and $\vec p_2$ in the decay plane, one obtains
\bea\label{Mphi}
&&\frac{1}{2\pi} \int_0^{2\pi} d\phi |{\cal M}_{R,L}|^2 \propto 
\nonumber \\ &&\qquad
|\vec J|^2(1 + \cos^2\theta) \pm 2{\rm Im}\left (\hat n\cdot (\vec J\times 
\vec J^*)\right )\cos\theta\,.
\eea
Using Eqs.~(\ref{A^2}) and (\ref{lambda}), one obtains the $B\to 
(K\pi\pi)_{K_1}\gamma$ decay distribution
\bea\label{distrib}
&& \frac{d\Gamma}{ds_{13}ds_{23}d\cos\theta} \propto
\nonumber \\ &&\qquad
|\vec J|^2(1 + \cos^2\theta)
+ \lambda_{\gamma} 2{\rm Im}\left (\hat n\cdot (\vec J\times 
\vec J^*)\right )\cos\theta\,.
\eea
Since the angular variable $\cos\theta$ changes sign 
under the exchange of $s_{13}$ and $s_{23}$, we define a new angle 
$\tilde\theta$ which is 
independent of $s_{13}$ and $s_{23}$, $\cos\theta \equiv 
\mbox{sgn}(s_{13}-s_{23})\,\cos\tilde\theta$. An equivalent definition of 
$\tilde\theta$ is the angle between $-\vec p_\gamma$ and the normal
to the decay plane defined by $\vec p_{\rm slow}\times \vec p_{\rm fast}$,
where  $\vec p_{\rm slow}$ and $\vec p_{\rm fast}$ are the momenta of the
slower and faster pions.

The asymmetry between decay distributions corresponding to right and
left-handed photons, from which the photon polarization can be determined,
is contained in the second term in Eq.~(\ref{distrib}). It describes an 
up-down
asymmetry of the photon momentum with respect to the $K_1$ decay plane.
In order to measure $\lambda_{\gamma}$ one would fit the $B$ and $\bar B$ decay
distributions to (\ref{distrib}), which has a well-defined dependence 
on $\theta$ and on the energy variables $s_{13},s_{23}$ occurring in the 
Breit-Wigner forms.
In order to obtain a conservative estimate for the sensitivity of the
decay distribution to the photon polarization, let us consider the 
integrated up-down asymmetry, 
\bea
{\cal A} &=& \frac{\int^{\pi/2}_0 \frac{d\Gamma}{d\cos\tilde\theta}
d\cos\tilde\theta
- \int^{\pi}_{\pi/2} \frac{d\Gamma}{d\cos\tilde\theta}d\cos\tilde\theta}
{\int^{\pi}_0 \frac{d\Gamma}{d\cos\tilde\theta}d\cos\tilde\theta}
\nonumber \\ 
 &=&
\frac34 \frac{\langle \mbox{Im}\,\left 
(\hat n\cdot (\vec J\times \vec J\,^*\right )~\mbox{sgn}(s_{13}-s_{23})
\rangle}{\langle |\vec J\,|^2\rangle}\, \lambda_\gamma\,.
\eea
Integrating the numerator and denominator over the entire Dalitz plot, one 
obtains 
\beq\label{calA}
{\cal A}=(0.34\pm 0.05)\lambda_\gamma~.
\eeq

The calculated asymmetry involves theoretical uncertainties 
from two sources: the $\rho K$ intermediate state which we neglected, and an 
error in the D-wave amplitude of $K_1 \to K^*\pi$. 
Varying the magnitude of the $K_1\rho K$ coupling under the constraint from 
the measured $K_1(1400)\to K\rho$ 
branching ratio, ${\cal B}(K_1\to K\rho) = 0.03 \pm 0.03$, 
and varying the relative intrinsic phase between the 
$\rho K$ and $K^*\pi$ amplitudes in the range $-30^{\circ}$ to $+30^{\circ}$  
as measured in \cite{Daum}, this amplitude introduces an uncertainty of $\pm 
0.04$ in ${\cal A}$. The experimental error in the $D$ wave amplitude 
is shown to contribute $\pm 0.03$ to this 
uncertainty when varying $|A_D/A_S|^2=0.04 \pm 0.01,~\delta_D - \delta_S = 
260^\circ\pm 20^\circ$ \cite{Daum}. 

The SM predicts $\lambda_\gamma \approx +1 (-1)$ for $B (\bar B)$ decays. 
Namely, {\sl in $B^-$ and $\bar B^0$ decays, 
the photon prefers to move in the hemisphere of $\vec p_{\rm 
slow}\times \vec p_{\rm fast}$, while in $B^+$ and $B^0$ decays it prefers 
to move in the opposite direction}.
For a three standard deviation measurement of a total up-down 
asymmetry, ${\cal A}~\simeq~0.34~(-0.34)$, expected in the SM for $B^+~(B^-)$ 
and $B^0~(\bar B^0)$ decays, one needs to observe a total of about  
80 charged and neutral $B$ and $\bar B$ decays to $(K\pi\pi)_{K_1}\gamma$. 
In order to estimate the number of $B\bar B$ pairs needed for this 
measurement, we will assume that the branching ratio of $B\to K_1(1400)\gamma$
is $0.7\times 10^{-5}$, as calculated in some models \cite{gK}. We use 
${\cal B}(K_1(1400) \to K^*\pi) = 0.94$ \cite{PDG}, and note 
that $4/9$ of all $K^*\pi$ events in $K_1^+$ and $K_1^0$ decays occur in
the two channels specified in Eq.~(\ref{chain}).
Including a factor $1/3$ for 
observing a $K_S$ (from $K^0$) through its $\pi^+\pi^-$ decay, we estimate 
a branching ratio of ${\cal B} = 0.7\times 10^{-5}\times (4/9)0.94  
\simeq 0.3\times 10^{-5}$ into $(K^+\pi^-\pi^0)_{K_1(1400)}$ and 
${\cal B} \simeq 0.1\times 10^{-5}$ into $(K_S\pi^+\pi^0)_{K_1(1400)}$.
Ignoring experimental efficiencies and background, 80 $(K\pi\pi)_{K_1}\gamma$
events can be obtained
from a total of $2\times 10^7$ $B \bar B$ pairs, including charged and 
neutrals. This number of $B$ mesons has already been produced at 
$e^+e^-$ colliders \cite{BaBar,BELLE,Cleo}. Since we ignored experimental 
efficiencies, resolution and background, one may have to wait a year or so 
before obtaining the required number of events.

Similar studies can be carried out for other kaon resonance 
states in radiative $B$ decays. The decay distribution for an excited $K^*_1$
is insensitive to the photon polarization.
In the case of $K^*_2(1430)$ one finds, when both $K^*\pi$ and $\rho K$ 
contributions are included,
\bea\label{distrib2}
& &\frac{d\Gamma}{ds_{13}ds_{23}d\cos\theta} = 
|\vec p_1\times \vec p_2|^2 \left[ |\vec J|^2(\cos^2\theta + \cos^2
2\theta) \right.\nonumber\\
& & + \left. \lambda_{\gamma} 2{\rm Im}\left(\hat n\cdot (\vec J\times 
\vec J^*)\right)
\cos\theta \cos 2\theta \right]~,
\eea
where
$\vec J  =  \vec p_1 [B^{K^*}_{23} + \kappa_\rho B^{\rho}_{12}] +
\vec p_2 [B^{K^*}_{13} + \kappa_\rho B^{\rho}_{12}]$ and 
$B^{\rho}_{12}$ is defined analogously to $B^{K^*}_{ij}$.
The complex parameter $\kappa_\rho$, parametrizing the relative
strength and final state phase difference of the $K^*\pi$ and $\rho K$ 
contributions, is given by
\bea
\kappa_\rho = |\kappa_\rho| e^{i\delta}
= \sqrt{\frac32} \frac{g_{K_2^* \rho K}}{g_{K_2^* K^* \pi}}\cdot
\frac{g_{\rho\pi\pi}}{ g_{K^* K\pi}} \simeq 2.38\,.
\eea
The two ratios of couplings are obtained from the corresponding measured
partial widths \cite{PDG}.
The strong phase $\delta$ vanishes in the SU(3) limit and is dominated by
the phase of $g_{K_2^* \rho K}/g_{K_2^* K^* \pi}$, measured to be smaller than 
$30^\circ$ in a $K^*_2$ resonance production experiment \cite{Daum}.

While the integrated up-down asymmetry in Eq.~(\ref{distrib2}) vanishes, 
a useful observable which is proportional to $\lambda_{\gamma}$ is 
$\langle\cos\tilde\theta \rangle$.
Integrating this quantity over a square region, $0.71 \mbox{GeV}^2 
\leq s_{13}, s_{23} \leq 0.89$ GeV$^2$, 
where the two $K^*$ bands of widths $2\Gamma_{K^*}$ overlap, 
one finds $\langle \cos\tilde\theta \rangle_s = (0.071 \pm 0.03)
\lambda_{\gamma}$, when $\delta$ is varied in the range $(0^\circ \pm 
30^\circ)$. The value of $\langle\cos\tilde\theta \rangle$ obtained when 
integrating over the entire Dalitz plot is considerably smaller. 

We conclude with a few practical comments. The region of $K\pi\pi$ invariant 
mass around 1400 MeV includes $K_1(1400)$ which involves the large up-down 
asymmetry calculated in (\ref{calA}), $K^*_1(1410)$ which leads to no 
asymmetry, and $K^*_2(1430)$ which adds a relatively small asymmetry. 
The two asymmetries from $K_1$ and $K^*_2$ have equal signs. Therefore,
the sign of the total asymmetry is predicted in the SM. Using the 
different energy and angular dependence of the three resonances, one
should be able to isolate the $K_1$ contribution from the other resonances
and from a small nonresonant $K\pi\pi$ contribution in a narrow invariant 
mass band around $m(K\pi\pi)=1400$ MeV. This would provide a first 
significant photon polarization measurement in radiative $b\to s\gamma$ decays,
which may confirm the SM prediction or detect a {\em large violation} of this 
prediction. A precise measurement,  
sensitive to {\em small} new physics effects, seems unfeasible at this time.
 
Our study focused on decay modes of higher $K$ resonances which involve one 
neutral pion. This was necessary in order to have two interfering $K^*\pi$ 
amplitudes which are related by isospin symmetry. 
An asymmetry is also expected in channels involving only charged particles, 
$K^{\pm}\pi^{\mp}\pi^{\pm}$, which were measured very recently by the Belle 
collaboration \cite{Belle}. In this case the asymmetry originates in the
interference between $K^*\pi$ and $\rho K$ (or $f_0K$) amplitudes. The latter
amplitude is significant in $K_1(1270)$ and $K^*_2(1430)$ decays. In
$K_1(1270)\to K^*\pi$ one must also consider the effect of a possibly 
significant $D$-wave amplitude, for which the upper limit is rather loose
\cite{PDG,Daum}. 

\medskip
We thank A. Weinstein for a useful correspondence and for sending us
his unpublished work \cite{Weinstein}.
D. P. is grateful to the Technion physics department, 
and M. G. wishes to thank the SLAC theory group for its kind hospitality.
This work was supported in part by the Israel Science Foundation
founded by the Israel Academy of Sciences and Humanities,
by the U. S. -- Israel Binational Science Foundation through Grant
No.\ 98-00237 and by the DOE under grant DOE-FG03-97ER40546.

\def \ajp#1#2#3{Am.\ J. Phys.\ {\bf#1} (#3) #2}
\def \apny#1#2#3{Ann.\ Phys.\ (N.Y.) {\bf#1} (#3) #2}
\def \app#1#2#3{Acta Phys.\ Polonica {\bf#1}, #2 (#3)}
\def \arnps#1#2#3{Ann.\ Rev.\ Nucl.\ Part.\ Sci.\ {\bf#1} (#3) #2}
\def \art{and references therein}
\def \cmts#1#2#3{Comments on Nucl.\ Part.\ Phys.\ {\bf#1} (#3) #2}
\def \cn{Collaboration}
\def \cp89{{\it CP Violation,} edited by C. Jarlskog (World Scientific,
Singapore, 1989)}
\def \efi{Enrico Fermi Institute Report No.\ }
\def \epjc#1#2#3{Eur.\ Phys.\ J. C {\bf#1}, #2 (#3)}
\def \f79{{\it Proceedings of the 1979 International Symposium on Lepton and
Photon Interactions at High Energies,} Fermilab, August 23-29, 1979, ed. by
T. B. W. Kirk and H. D. I. Abarbanel (Fermi National Accelerator Laboratory,
Batavia, IL, 1979}
\def \hb87{{\it Proceeding of the 1987 International Symposium on Lepton and
Photon Interactions at High Energies,} Hamburg, 1987, ed. by W. Bartel
and R. R\"uckl (Nucl.\ Phys.\ B, Proc.\ Suppl., vol.\ 3) (North-Holland,
Amsterdam, 1988)}
\def \ib{{\it ibid.}~}
\def \ibj#1#2#3{~{\bf#1} (#3) #2}
\def \ichep72{{\it Proceedings of the XVI International Conference on High
Energy Physics}, Chicago and Batavia, Illinois, Sept. 6 -- 13, 1972,
edited by J. D. Jackson, A. Roberts, and R. Donaldson (Fermilab, Batavia,
IL, 1972)}
\def \ijmpa#1#2#3{Int.\ J.\ Mod.\ Phys.\ A {\bf#1} (#3) #2}
\def \ite{{\it et al.}}
\def \jhep#1#2#3{JHEP {\bf#1}, #2 (#3)}
\def \jpb#1#2#3{J.\ Phys.\ B {\bf#1} (#3) #2}
\def \lg{{\it Proceedings of the XIXth International Symposium on
Lepton and Photon Interactions,} Stanford, California, August 9--14 1999,
edited by J. Jaros and M. Peskin (World Scientific, Singapore, 2000)}
\def \lkl87{{\it Selected Topics in Electroweak Interactions} (Proceedings of
the Second Lake Louise Institute on New Frontiers in Particle Physics, 15 --
21 February, 1987), edited by J. M. Cameron \ite~(World Scientific, Singapore,
1987)}
\def \kdvs#1#2#3{{Kong.\ Danske Vid.\ Selsk., Matt-fys.\ Medd.} {\bf #1},
No.\ #2 (#3)}
\def \ky85{{\it Proceedings of the International Symposium on Lepton and
Photon Interactions at High Energy,} Kyoto, Aug.~19-24, 1985, edited by M.
Konuma and K. Takahashi (Kyoto Univ., Kyoto, 1985)}
\def \mpla#1#2#3{Mod.\ Phys.\ Lett.\ A {\bf#1} (#3) #2}
\def \nat#1#2#3{Nature {\bf#1} (#3) #2}
\def \nc#1#2#3{Nuovo Cim.\ {\bf#1} (#3) #2}
\def \nima#1#2#3{Nucl.\ Instr.\ Meth. A {\bf#1} (#3) #2}
\def \npb#1#2#3{Nucl.\ Phys.\ B~{\bf#1}, #2 (#3)}
\def \os{XXX International Conference on High Energy Physics, 27 July
-- 2 August 2000, Osaka, Japan}
\def \PDG{Particle Data Group, D. E. Groom \ite, \epjc{15}{1}{2000}}
\def \pisma#1#2#3#4{Pis'ma Zh.\ Eksp.\ Teor.\ Fiz.\ {\bf#1} (#3) #2 [JETP
Lett.\ {\bf#1} (#3) #4]}
\def \pl#1#2#3{Phys.\ Lett.\ {\bf#1} (#3) #2}
\def \pla#1#2#3{Phys.\ Lett.\ A {\bf#1} (#3) #2}
\def \plb#1#2#3{Phys.\ Lett.\ B {\bf#1}, #2 (#3)}
\def \pr#1#2#3{Phys.\ Rev.\ {\bf#1} (#3) #2}
\def \prc#1#2#3{Phys.\ Rev.\ C {\bf#1} (#3) #2}
\def \prd#1#2#3{Phys.\ Rev.\ D {\bf#1}, #2 (#3)}
\def \prl#1#2#3{Phys.\ Rev.\ Lett.\ {\bf#1}, #2 (#3)}
\def \prp#1#2#3{Phys.\ Rep.\ {\bf#1} (#3) #2}
\def \ptp#1#2#3{Prog.\ Theor.\ Phys.\ {\bf#1} (#3) #2}
\def \rmp#1#2#3{Rev.\ Mod.\ Phys.\ {\bf#1} (#3) #2}
\def \rp#1{~~~~~\ldots\ldots{\rm rp~}{#1}~~~~~}
\def \si90{25th International Conference on High Energy Physics, Singapore,
Aug. 2-8, 1990}
\def \slc87{{\it Proceedings of the Salt Lake City Meeting} (Division of
Particles and Fields, American Physical Society, Salt Lake City, Utah, 1987),
ed. by C. DeTar and J. S. Ball (World Scientific, Singapore, 1987)}
\def \slac89{{\it Proceedings of the XIVth International Symposium on
Lepton and Photon Interactions,} Stanford, California, 1989, edited by M.
Riordan (World Scientific, Singapore, 1990)}
\def \smass82{{\it Proceedings of the 1982 DPF Summer Study on Elementary
Particle Physics and Future Facilities}, Snowmass, Colorado, edited by R.
Donaldson, R. Gustafson, and F. Paige (World Scientific, Singapore, 1982)}
\def \smass90{{\it Research Directions for the Decade} (Proceedings of the
1990 Summer Study on High Energy Physics, June 25--July 13, Snowmass,
Colorado),
edited by E. L. Berger (World Scientific, Singapore, 1992)}
\def \tasi{{\it Testing the Standard Model} (Proceedings of the 1990
Theoretical Advanced Study Institute in Elementary Particle Physics, Boulder,
Colorado, 3--27 June, 1990), edited by M. Cveti\v{c} and P. Langacker
(World Scientific, Singapore, 1991)}
\def \yaf#1#2#3#4{Yad.\ Fiz.\ {\bf#1} (#3) #2 [Sov.\ J.\ Nucl.\ Phys.\
{\bf #1} (#3) #4]}
\def \zhetf#1#2#3#4#5#6{Zh.\ Eksp.\ Teor.\ Fiz.\ {\bf #1} (#3) #2 [Sov.\
Phys.\ - JETP {\bf #4} (#6) #5]}
\def \zpc#1#2#3{Zeit.\ Phys.\ C {\bf#1}, #2 (#3)}
\def \zpd#1#2#3{Zeit.\ Phys.\ D {\bf#1} (#3) #2}

\end{document}